\documentclass[twocolumn,aps,prc,showpacs]{revtex4}
\usepackage{epsfig}
\usepackage{graphicx}
\begin{document}

\title{Contribution to the study of narrow low mass hadronic structures}

\author{B. Tatischeff}
\affiliation{\it CNRS/IN2P3, Institut de Physique Nucl\'eaire, UMR 8608, 
\\ and  Univ. Paris-Sud, Orsay, 91405 Orsay, France}
\thanks{e-mail : tati@ipno.in2p3.fr}

\author{E. Tomasi-Gustafsson}
\affiliation{\it DAPNIA/SPhN, CEA/Saclay, 91191 Gif-sur-Yvette Cedex,
France }


\vspace*{1cm}
\begin{abstract}
New data are presented, concerning narrow exotic structures in mesons, baryons and dibaryons. The sequence of narrow baryons is quite well described starting from the sequence of narrow mesons. In the same way, the sequence of narrow dibaryons is rather well described starting from the sequence of narrow baryons. Lastly it is shown that the masses of these narrow hadronic structures lie on straight line Regge-like trajectories.  
\end{abstract}
\maketitle
\section{Introduction}
This paper presents new data concerning narrow hadronic structures in several species, namely in mesons, baryons, and dibaryons. Many data were already published, and therefore are not repeated here, the corresponding papers are only quoted. Some new data, not published up to now, are shown. 

It is also desirable to look at possible similitudes between mass spectra of the three hadronic species. Indeed, if such connection exist, it will give a strong argument in favor of the genuine existence of these structures, which therefore could not be associated to accidental statistical effects. 
\section{Mesons}
\subsection{Short summary of previously published data}
A paper, recently published \cite{btetg}, shows evidence for narrow and weakly excited mesonic structures, with masses below and just above the pion mass (M=139.56~MeV). These data are mainly missing mass precise spectra of the  pp$\to$ppX reaction, studied at SPES3 (Saturne) and also selected results from COSY, Celsius, MAMI, and JLAB Hall A, Hall B,  and Hall C. The statistical confidence is often large. In some cases, this confidence is not large, but several structures at approximately the same masses are observed.
These masses are M=62~MeV, 80~MeV, 100~MeV, 181~MeV, 198~MeV, 215~MeV, 227.5~MeV, and 235~MeV, although the last one may be uncertain, since determined by only three data, and being located at the limit of the spectra. 

Several papers \cite{jy} \cite{bt1} show, some time ago, evidence for narrow mesonic structures at masses larger than the pion mass. These masses are: M=310~MeV, 350~MeV, 430~MeV, (495)~MeV, 555~MeV, 588~MeV, 608~MeV, 647~MeV, 681~MeV, 700~MeV, and 750~MeV.\\
\subsection{New data from the $\gamma$p$\to$p$\pi^{+}\pi^{-}$ reaction studied at MAMI}
\begin{center}       
\begin{figure}
\scalebox{.48}[.35]{
\includegraphics[bb=10 10 550 550,clip=]{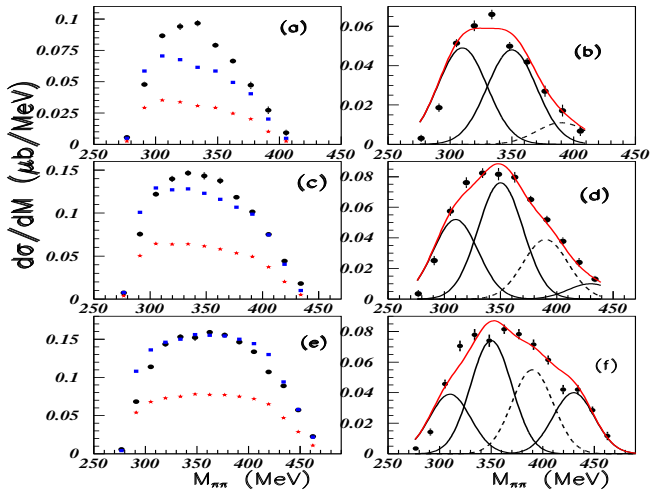}}
\caption{Tentative reanalysis of the $\gamma$p$\to$p$\pi^{+}\pi^{-}$ reaction studied at MAMI \protect\cite{ahrens}. Inserts (a), (b), and (c) show the data (full circles), the values of the FA model \protect\cite{arenhovel} (full squares), and the same after renormalization (stars).
Inserts (b), (d), and (f) show the differences and their analysis into fixed mass gaussians. Inserts (a) and (b) correspond to 500$\le~T_{\gamma}\le$550~MeV, inserts (c) and (d) correspond to 
550$\le~T_{\gamma}\le$600~MeV, and inserts (e) and (f) correspond to 
600$\le~T_{\gamma}\le$650~MeV.}
\label{Fig1}
\end{figure}

\begin{figure}
\scalebox{.48}[.35]{
\includegraphics[bb=10 10 550 550,clip=]{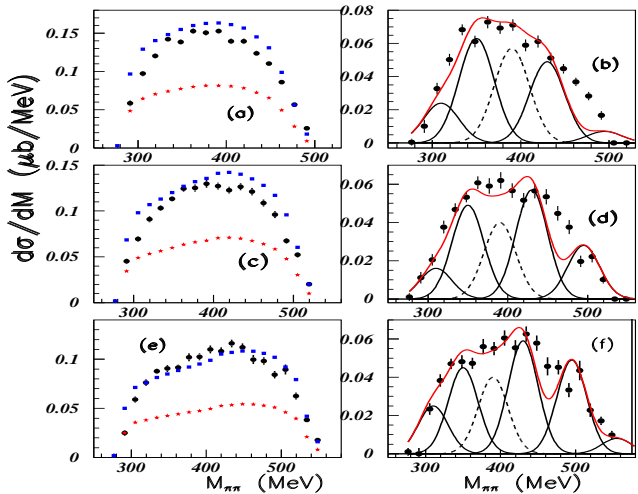}}
\caption{Same caption as for Fig.~1, except that inserts (a) and (b) correspond to 650$\le~T_{\gamma}\le$700~MeV, inserts (c) and (d) correspond to 
700$\le~T_{\gamma}\le$750~MeV, and inserts (e) and (f) correspond to 
750$\le~T_{\gamma}\le$800~MeV.}
\label{Fig2}
\end{figure}
\end{center}
The helicity dependence for the $\gamma$p$\to$p$\pi^{+}\pi^{-}$ reaction was studied with the detector DAPHNE at the tagged photon beam facility at the MAMI accelerator in Mainz \cite{ahrens}. The cross sections were assembled into six ranges of photon incident energy, with photon energy interval widths of 50~MeV, altogether between 500~MeV and 800~MeV.
The spectra were compared by the authors \cite{ahrens} to the predictions of the Fix-Arenh\"{o}vel model \cite{arenhovel} (FA).These spectra, versus the two-pion invariant mass M$_{\pi\pi}$, show some small structures well outside the statistical uncertainties. 

An attempt is done here, to look how better agreement can be achieved, with the assumption to add small narrow mesonic structures to the Fix-Arenh\"{o}vel model, at the same masses as those extracted from previous papers. Since the model gives in some cases  larger cross sections than the measured ones, we have arbitrarily renormalized the model values by 0.5, in order to suppress all negative differences. Fig.~1 and fig.~2 show the cross sections for the six incident photon energies. The masses of the gaussians are M$_{\pi\pi}$=310~MeV, 350~MeV,
390~MeV, 430~MeV, 495~MeV, and 555~MeV. A gaussian at M=390~MeV is introduced here although not extracted before. The gaussians widths are fixed to $\sigma$=20~MeV. Fig.~3, and fig.~4 show the cross sections for the spin-1/2 state for the p$\pi^{\pm}$ system. 
\begin{figure}[t]
                                                
\scalebox{.48}[.35]{
\includegraphics[bb=10 10 550 550,clip=]{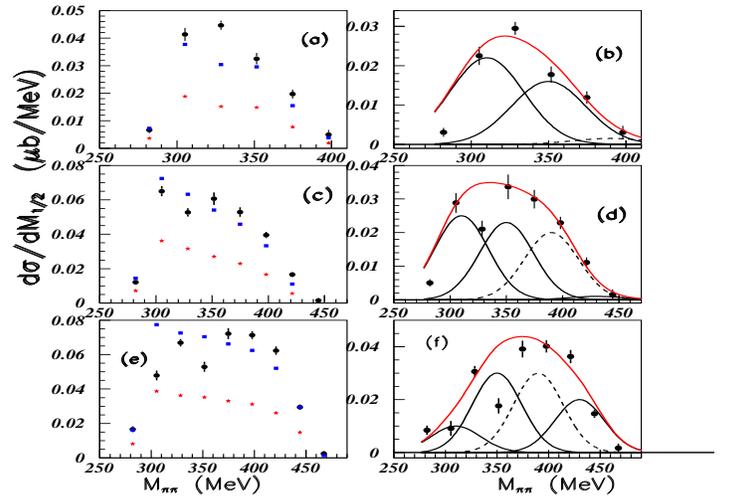}}
\caption{Same caption as for fig.~1, except that here the cross sections are for the spin-1/2 state for the p$\pi^{\pm}$ system}
\label{fig3}

\end{figure}
\begin{figure}[t]
                                                        
\scalebox{.48}[.45]{
\includegraphics[bb=10 10 550 550,clip=]{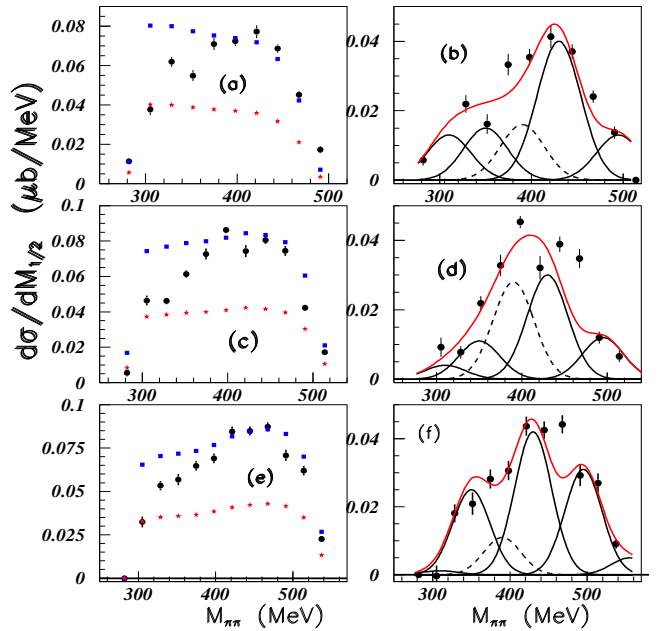}}
\caption{Same caption as for fig.~2, except that here the cross sections are for the spin-1/2 state for the p$\pi^{\pm}$ system}
\label{fig4}

\end{figure}
Fig.~5 and fig.~6 show the cross sections for the spin-3/2 state for the p$\pi^{\pm}$ system. The same masses are used for all six figs., but in figs.~3, 4, 5, and 6 a larger width ($\sigma$=24~MeV) is used for gaussians, since the experimental binning is larger.
\begin{figure}[b]
                                                       
\scalebox{.48}[.35]{
\includegraphics[bb=10 10 550 550,clip=]{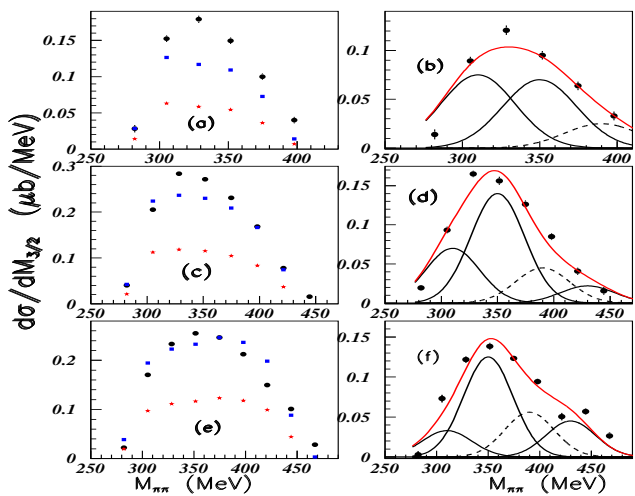}}
\caption{Same caption as for fig.~1, except that here the cross sections are for the spin-3/2 state for the p$\pi^{\pm}$ system.}
\label{fig5}

\end{figure}
\begin{figure}[t]
                                                       
\scalebox{.48}[.35]{
\includegraphics[bb=10 10 550 550,clip=]{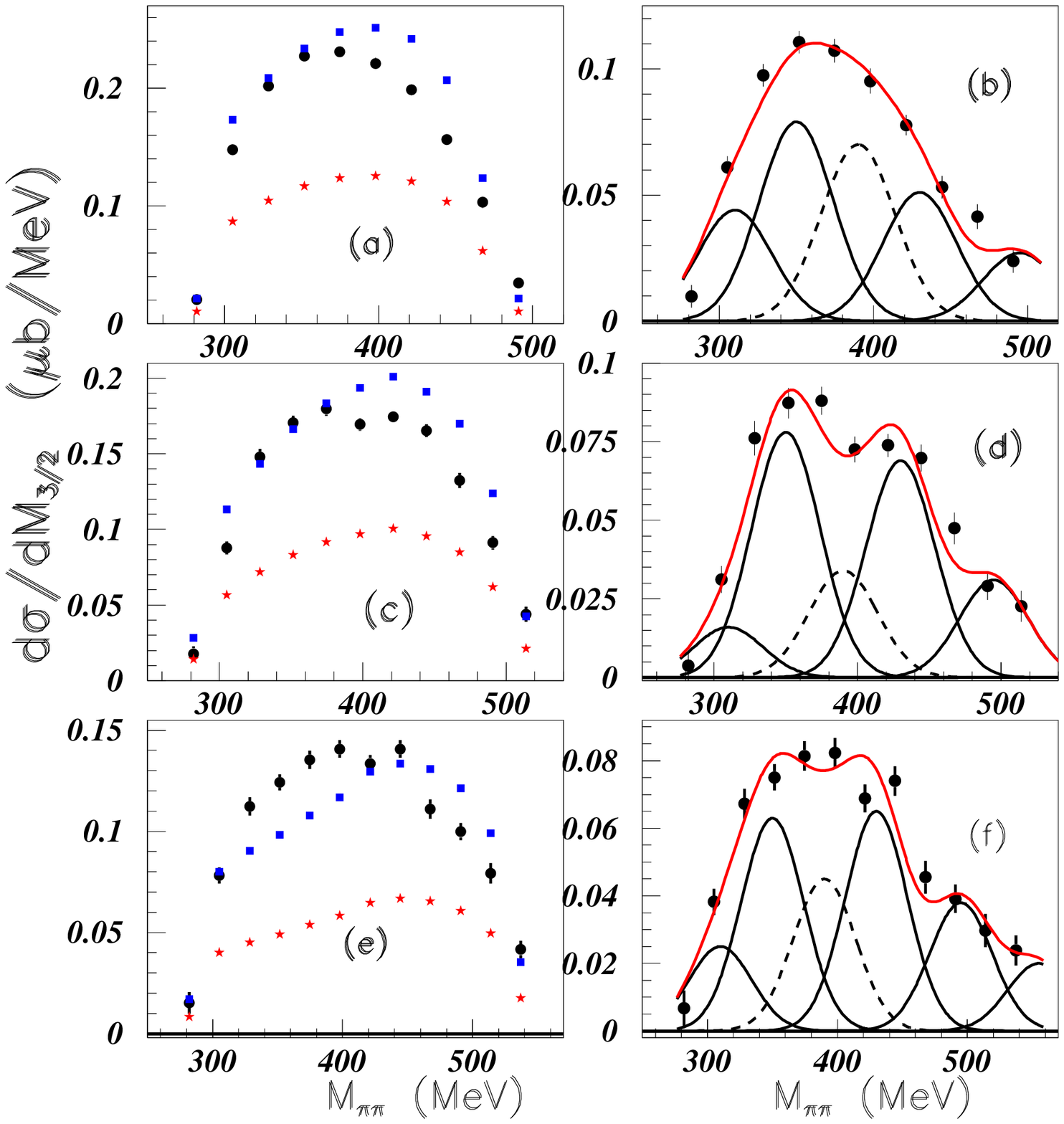}}
\caption{Same caption as for fig.~2, except that here the cross sections are for the spin-3/2 state for the p$\pi^{\pm}$ system.}
\label{fig6}

\end{figure}
\begin{figure}[!b]
                                                       
\scalebox{.48}[.35]{
\includegraphics[bb=10 10 550 550,clip=]{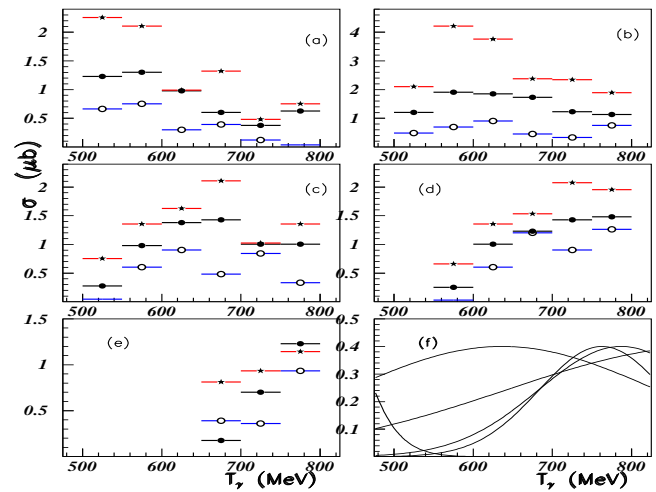}}
\caption{Total cross sections of the mesonic narrow structures. The five inserts: (a), (b), (c), (d), and (e), correspond respectively to the following mesonic structure masses: M=310, 350, 390, 430, and 495~MeV. The full circles show the cross-section, the stars show the cross section for spin-3/2 state for the p$\pi^{\pm}$ system., and the empty circles show the same for spin-1/2 states. The last insert (f), shows the display of the broad PDG baryonic resonances arbitrarily normalized.}
\label{fig7}
\end{figure}
Fig.~7 shows the integral of the narrow structure cross sections. We observe that each insert, which corresponds to a specific narrow mesonic structure mass, show a regular shape, smoothly varying with the incident photon energy. We observe that the M=310~MeV structure is mainly influenced by the $\Delta(3/2,3/2)$(1232) resonance.
The last insert (f) shows, the display of the broad PDG baryonic resonances arbitrarily normalized.
We observe that the shapes of all narrow mesonic increasing mass structures, are "influenced" by increasing PDG baryonic resonance masses. However such property cannot help us to suggest a possible isospin or spin, to the narrow mesonic structures.
\subsection{The old pp$\to$dX data}
The  pp$\to$dX reaction was studied in Birmingham \cite{hall} long time ago. In order to study the phase shifts of the T=0 $\pi - \pi$ interaction, the authors compare the correponding cross sections with those obtained using a deuteron target. They extracted the spectrum of the T=0 $\to$ d2$\pi$ which was plotted versus p$_{d}$, the detected deuteron momentum (their fig. 4). This spectrum was read, and reported in fig.~8(a) versus the missing mass of the reaction. We observe the two branches of the kinematics at both sides of the maximum missing mass M$_{X}$=430~MeV. 
Indeed the incident proton energy T$_{p}$=991~MeV, and the deuteron lab. scattering angle $\theta_{d}$=4.2$^{0}$ allow a maximum missing mass M$_{X}$=430~MeV. In fig.~8(a)
we reported below the data, their calculation of the enhanced phase space incorporating a value of 1.0 
$\hbar/mc$ 
for the $\pi\pi$ S-wave scattering length, renormalized by 0.5 in order to be lower than the data in all spectrum range.

\begin{figure}
\scalebox{.48}[.35]{
\includegraphics[bb=10 10 550 550,clip=]{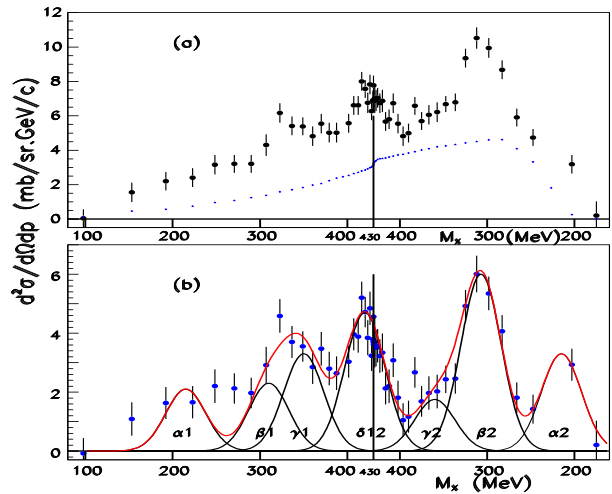}}
\caption{Spectrum of the T=0 pn$\to$dX reaction \protect\cite{hall}. Insert (a) shows the data plotted versus the missing mass M$_{X}$. Insert (b) shows the narrow mesonic structures extracted from the data minus phase space (see text).}
\label{fig8}

\end{figure}
Fig.~8(b) shows the data subtracted from this phase space. Clear structures are extracted, which are already seen in insert (a). The real shape of the space space is therefore not important, provided it does not present narrow structures. Table I shows that the masses of the extracted peaks reproduced fairly well the narrow structure masses extracted previously from the SPES3 data. 
\begin{table}[h]
\vspace{5.mm}
\begin{tabular}{c c c c c c c c c c c}
\hline
&$\alpha$&&&$\beta$&&&$\gamma$&&&$\delta$\\
\hline
1&2&spes3&1&2&spes3&1&2&spes3&12&spes3\\
215&215&215&310&307&310&350&360&350&420&430\\
\hline
\end{tabular}
\caption{Masses (in MeV) of the forward (1) and backward (2) pionic missing mass branches, from the pN$\to$dX reaction \protect\cite{hall}. In comparison the masses extracted from SPES3 experiments are also given.}
\label{Table I}
\end{table}
All structures are extracted with gaussians with an unique width $\sigma$=23~MeV.
\begin{figure}[t]
                                                       
\scalebox{.48}[.35]{
\includegraphics[bb=10 10 550 550,clip=]{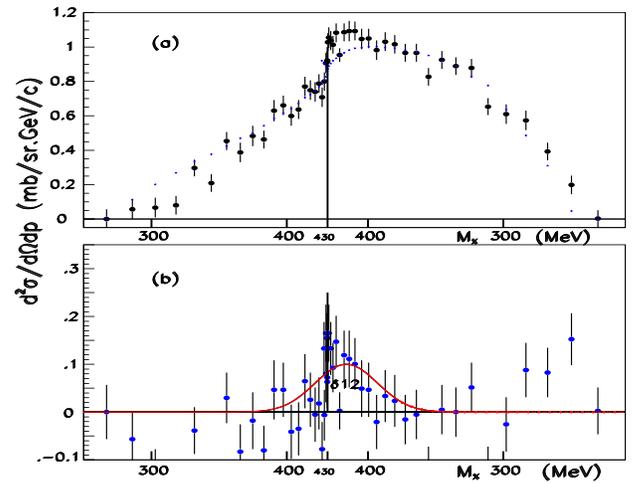}}
\caption{Same caption as for fig.~8, except that here the cross sections are for the isospin T=1.}
\label{fig9}
\end{figure}
Fig.~9 shows the corresonding results for T=1 isospin state (pp$\to$dX reaction). Here, except a possible very small structure close to M$_{X}$=430~MeV, no clear isevectorial structure can be extracted. This 
agrees with the Bose-Einstein statistics. Indeed these low mass pionic structures have l=0 orbital momenta, and the total two-pion isospin must be even. This agrees also with the spin and isospin values predicted for these states from the SPES3 data \cite{jy}.
\subsection{Discussion}
Several other data, sometimes not very precise, exist, which exhibit the presence of structures not belonging to classical PDG mesons. For example, the missing mass of the pp$\to$ppX reaction, measured at Argonne National Laboratory \cite{anderson}. In their fig.~3(b) a clear peak is observed between the $\pi^{0}$ and the $\eta$ mesons. An additionnal argument in favor of these structures can be put forward, if the corresponding masses display a continuous behaviour when displayed on Regge trajectories. 

It is accepted that all hadrons are Reggeons, that is lie on Regge trajectories. This was studied in \cite{btetg} where it was shown that three straight lines describe the Regge-like trajectories.  Fig.~10 reproduces this result. 
$$ NJ=a + b.M^{2}\hspace*{3.cm} [1]$$

We name them Regge-like, since the integer values in the ordinate - NJ -  represent an unknown quantum number. The first line lies below one pion mass, the second line lies between one and two pion masses, and the third line lies above two pion masses. We get very large slopes "b" (in GeV$^{-2}$) - see table II, when the corresponding value from the broad PDG mesons is close to b=0.877 \cite{afonin}.
\begin{figure}[h]
                                                       
\scalebox{.48}[.35]{
\includegraphics[bb=20 20 530 530,clip=]{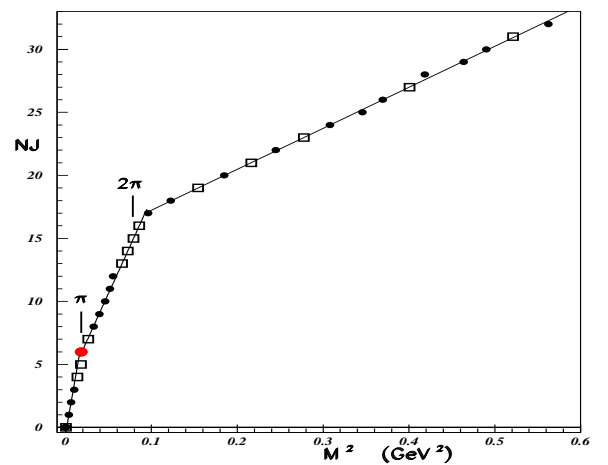}}
\caption{Regge-like trajectory of all narrow structure mesonic masses observed in this work and in \protect\cite{jy} \protect\cite{bt1}.}
\label{fig10}

\end{figure}
 In previous works showing evidence for narrow hadronic structures, their masses were computed using a mass formula for two clusters of quarks at the
ends of a stretched bag derived some years ago in terms of color
magnetic interactions \cite{mul}: 
\vspace*{0.mm}
$$ M=M_0+M_1[i_1(i_1+1)+i_2(i_2+1)+(1/3)$$
\vspace*{-1.1cm}

$$\hspace{1.3cm}(s_1(s_1+1)+s_2(s_2+1))]\hspace*{2.cm} [2]$$

where $M_0$ and $M_1$ are parameters deduced in \cite{mul} from mesonic and baryonic mass spectra, and $i_1(i_2)$, $s_1(s_2)$ are isospin, spin of the first (second)
quark cluster. Although it is not proven that these structures are a manifestation of colored 
quark clusters, we have considered such assumption and shown that the use of such mass formula, with the condition to be phenomenological, allows to reproduce the experimentally observed masses with only two adjustable parameters. Notice that the formula give rise to degenerascy in spin and isospin.

Now, using the Regge-like trajectory slopes, we are able to suggest the masses of both clusters. We use a simple relation, derived long time ago by Barut \cite{barut} \cite{ackers}, which relates (with several assumptions) the slope "b" of the Regge-like trajectories with the masses of two interacting clusters:

$$ \lambda=b^{-1} = 8 m_{1}m_{2}/137 \hspace*{3.cm}[3]$$

where m$_{1}$ and m$_{2}$ are the masses of both clusters. Using the assumption that the mass of the constituent quarks m$_{q}$ = m$_{\overline q}$ is close to 310~MeV, the previous formula allows us to determine the clusters through the calculation of the corresponding quark mass. The results of such calculations for mesons are shown in
table~II. The clusters giving a mass close to M=310~MeV are favoured.

We observe that the ($q{\overline q}$)-($q{\overline q}$) clusters are the most appropriate for the present analysis. For the larger masses studied here, the $q^{3}-{\overline q}^{3}$ clusters may also be appropriate. For mesons, we have made the assumption that the clusters are  
(q${\overline q})^{2}$. The $q-{\overline q}$ configurations are not introduced, since there is no room for such configurations 
in the field of narrow exotic mesonic structures. For these low mass mesonic structures, we have l=0 orbital momenta between both quark clusters, and then all levels have positive parities.

It was previously shown, in \cite{jy} \cite{bt1}, that the experimentally observed narrow  mesonic structure masses were satisfactory reproduced above two pion threshold mass, by calculation, using mass formula [1], with M$_{0}$=310~MeV and M$_{1}$=30~MeV.
Table~II shows that above two-pion threshold mass, where the Regge-like trajectory slope
equals 32.5~GeV$^{-2}$, the quark cluster configurations should be first ($q{\overline q}$)-($q{\overline q}$), and for larger masses $q^{3}-{\overline q}^{3}$. This was 
indeed done in \cite{jy}, where above M=470~MeV the masses were calculated using 
$q^{3}-{\overline q}^{3}$ configurations, and above M=620~MeV using 
$q^{4}-{\overline q}^{4}$ configurations \cite{bt1}.

Here, below two-pion mass, because the large value of the Regge-like trajectory slope, the
($q{\overline q}$)-($q{\overline q}$) configurations must be used. This is done in fig.~11.
There is no obligation to use the same values of both parameters, as those used in the mass range above two pion threshold. We keep however the same value of M$_{1}$=30~MeV, and adjust only M$_{0}$=60~MeV, in order to get the masses shown in fig.~11.
\begin{figure}[!h]
                                                       
\scalebox{.48}[.35]{
\includegraphics[bb=20 100 530 530,clip=]{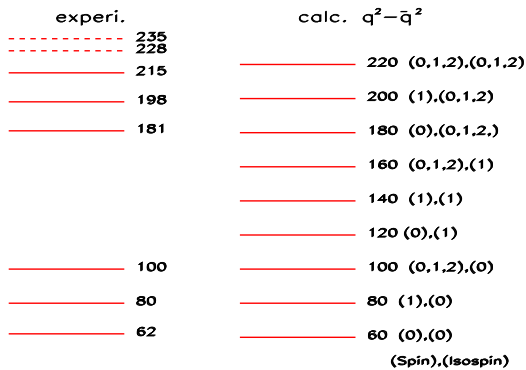}}
\caption{Experimental and calculated masses of narrow low mass mesonic structures.
The quantum numbers for calculated levels are (S),(I).}
\label{fig11}
\end{figure}

\begin{table}[h]
\vspace{5.mm}
\begin{tabular}{c c c c c}
\hline
b (GeV$^{-2}$)&($q{\overline q}-q{\overline q}$)&($q^{3}-\overline q^{3})$&M$_{min}$&M$_{max}$\\
390&104.8&69.8&0&135\\
149.7&169.1&112.7&235&300\\
32.5&{\bf 362.9}&{\bf 242}&300&760\\
\hline
\end{tabular}
\caption{Constituent quark mass, for different clusters supposed to describe the mesonic narrow structures, derived from the Barut's equation \protect\cite{barut}. M$_{min}$ and M$_{max}$ give, in MeV, the mass range of each trajectory.}
\label{Table II}
\end{table}
If we omit the three predicted structures at M=120~MeV, 140~MeV, and 160~MeV, we observe the agreement between experimental and calculated masses. The calculated mass at M=140~MeV cannot be the pion, since the spin here is J=1. More, the large pion peak excitation is clearly related to $q\overline{q}$ configurations, These three predicted small structures at M=120~MeV, 140~MeV, and 160~MeV, could only be looked for in dedicated experiments with very good resolution, therefore at low incident energy and forward angle missing mass experiments.
\section{Baryons}
\subsection{Previously shown narrow baryonic structures data}
Narrow baryonic structures were observed, first in the mass range 1.0$\le$M$\le$1.4~GeV 
studied at SPES3 (Saturne) with use of the pp$\to$p$\pi^{+}$X reactions \cite{bo1}.
This study was followed by the introduction of the data from the pp$\to$ppX reaction
 \cite{bo2} and then confirmed by a careful scrutiny of different cross-sections obtained as well with incident hadrons as with incident leptons. The extracted data were reported in several papers, corresponding to the following ranges: 1.47$\le$M$\le$1.68~GeV \cite{bo4},
172$\le$M$\le$1.79~GeV \cite{bo3}. The masses in the mass range 1.0$\le$M$\le$1.14~GeV are: M=1.004, 1044, 1094, 1136, 1173, 1249, 1277, 1339, and 1384~MeV. In the mass range 1.46$\le$M$\le$1.68~GeV, they are: M=1479, 1505, 1517, 1533, 1542, 1554, 1564, 1577, 1601, 1622, 1639, 1659, and 1669~MeV.  In the mass range 1.72$\le$M$\le$1.79~GeV, they are: M=1747 and 1772~MeV.
\subsection{New data from the missing mass of the pp$\to$p$\pi^{+}$X reaction studied at SPES3 (Saturne)}
The low mass range M$_{N}\le$M$\le$1.0~GeV was also studied, but only preliminary results \cite{bo5} were shown up to now. 
\begin{figure}[!t]
\scalebox{.49}[.35]{
\includegraphics[bb=40 360 565 812,clip=]{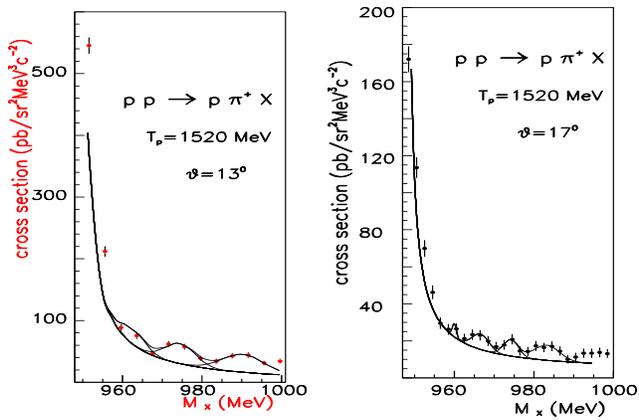}}
\caption{Selection of two cross-sections at T$_{p}$=1520~MeV, showing an oscillatory pattern in the missing mass.}
\label{fig12}
\end{figure}
\begin{figure}[!h]
\scalebox{.49}[.35]{
\includegraphics[bb=60 300 570 750,clip=]{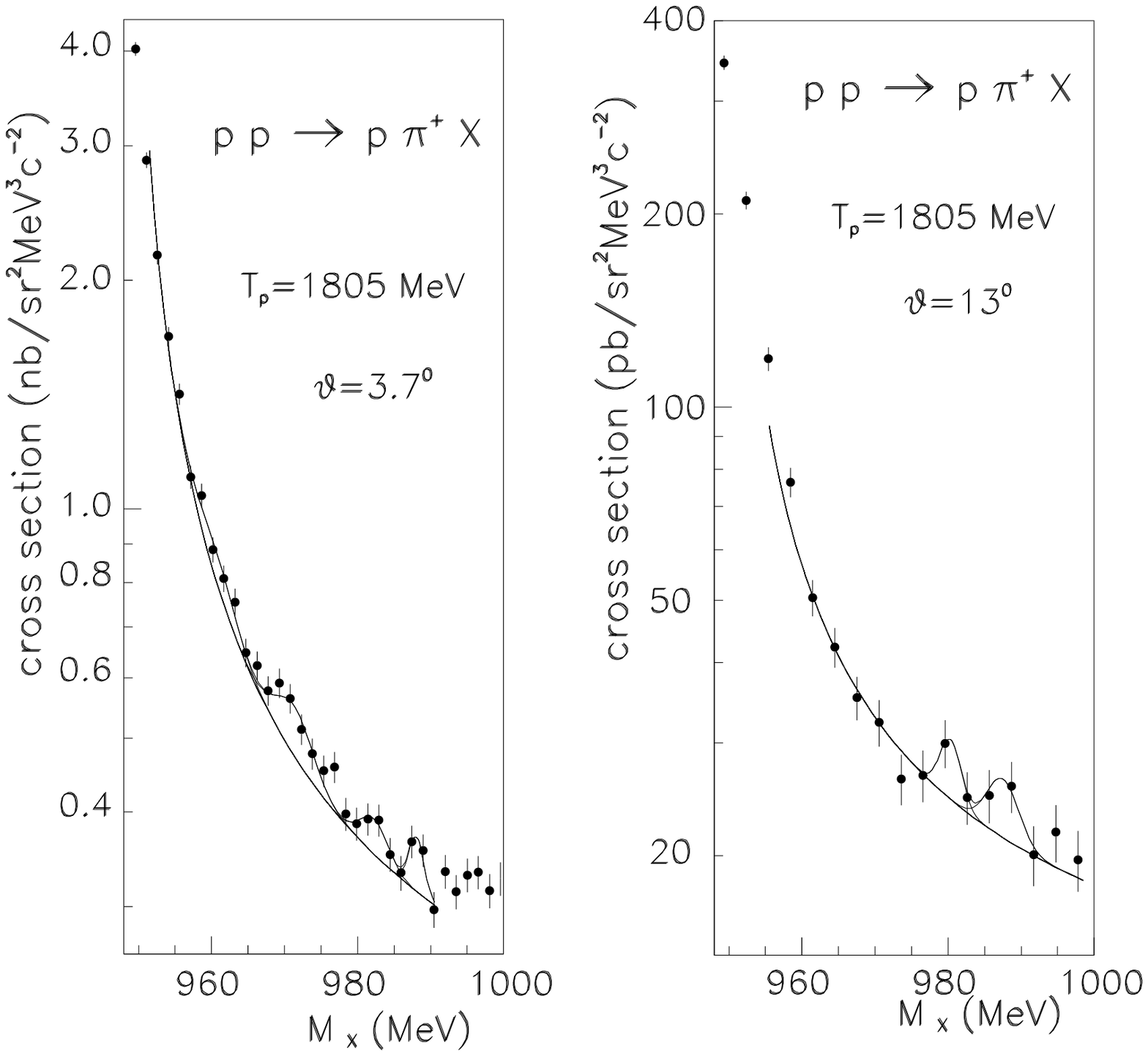}}
\caption{Selection of two cross-sections at T$_{p}$=1805~MeV, showing an oscillatory pattern in the missing mass.}
\label{fig13}

\end{figure}

\begin{figure}[t]
                                                       
\scalebox{.49}[.35]{
\includegraphics[bb=60 300 550 750,clip=]{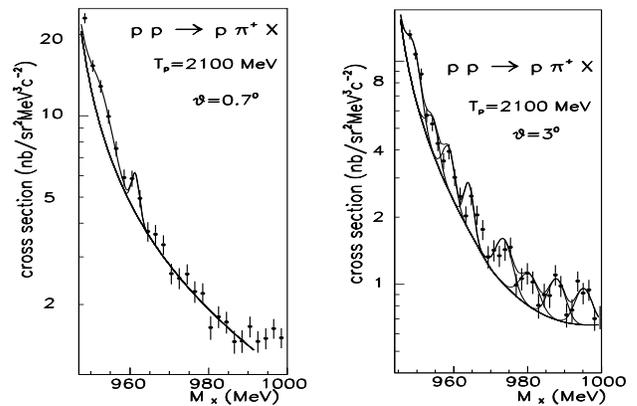}}
\caption{Selection of two cross-sections at T$_{p}$=2100~MeV, showing an oscillatory pattern in the missing mass.}
\label{fig14}

\end{figure}

Therefore the results in this low M$_{N}\le$M$\le$1.0~GeV mass range are presented here.
Figs. 12, 13, and 14 show a selection of cross-sections from pp$\to$p$\pi^{+}$X reactions studied at SPES3 (Saturne).  
Each fig. shows two spectra at the same incident energy and two different spectrometer angles. 
We observe "oscillatory paterns", allowing an attempt to define narrow structure masses.
Here the background determination is
clearly somewhat ambiguous. Several different choices can be used for
background. The first
one consists to draw an averaged background inside the data. Then all structures
will be strongly reduced, if not disappear. However we consider as quit unlikely
the situation where many spectra will exhibit an oscillatory pattern
and where all these oscillations will be accidental. The second choice consists
to consider these variations as physical, and to give them the same width as the
experimental neutron missing mass widths which increased slowly with the
spectrometer angle. Then the
structures will be extracted with a higher number of
standard deviation (S.D.). We choose the third - intermediate - choice and
draw the background using the low data points as in Fig. 12
Although 54 peaks were extracted from all cross
section spectra, only 18 were kept which had a
valuable statistical significance, since they were extracted with
S.D.$\ge$~3. The corresponding masses are displayed
in Fig. 15. When a mass ($\pm$~1~MeV) is obtained at least twice, it is
considered as being a candidate for a new structure and an horizontal dashed
range is drawn in Fig. 15. These masses are: M=950~MeV, 955.4~MeV, 961.5~MeV, 973~MeV, 988~MeV, and 994.8~MeV. The figure shows also the two narrow masses (at 966 MeV and
986 MeV), extracted by L.V. Filkov {\it et al.} from the
p~d~$\rightarrow$~p~p~X$_{1}$ reaction \cite{fil}.
\begin{figure}[!h]
                                                       
\scalebox{.49}[.35]{
\includegraphics[bb=20 20 550 550,clip=]{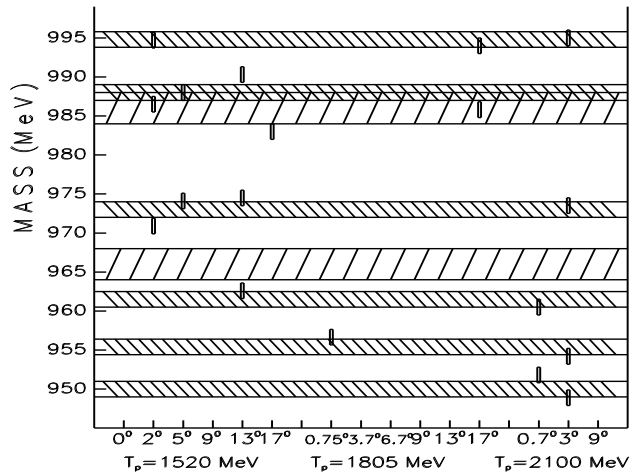}}
\caption{Low narrow exotic baryonic masses.}
\label{fig15}

\end{figure}
The spectra showing the narrow baryonic structures at larger masses, will not be shown here, since they were reported in previous publications \cite{bo1,bo2,bo3,bo4}.

\subsection{Discussion}
As done above for mesons, the study of Regge-like trajectories of the narrow structure baryonic masses is done and shown in fig.~16.
\begin{figure}[b]
                                                       
\scalebox{.48}[.40]{
\includegraphics[bb=20 20 530 530,clip=]{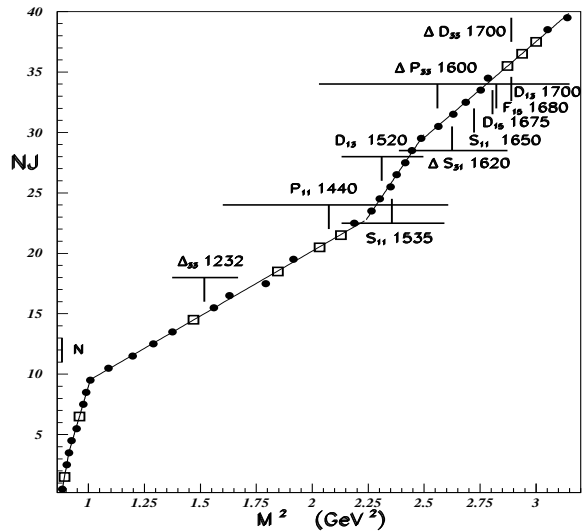}}
\caption{Regge-like trajectory of all narrow structure baryonic masses observed in this work and in all other references quoted in the text.}
\label{fig16}

\end{figure}
Here five different slopes are observed: the first two slopes: b=100.4~GeV$^{-2}$
 and b=62.6~GeV$^{-2}$ in the region close to N, the third one: b=10.7~GeV$^{-2}$ in the mass region of the first $\Delta$(3/2, 3/2), then the next slope b=27.3~GeV$^{-2}$ in the mass region of several N$^{*}$, and finally 
b=15.9~GeV$^{-2}$  in the next mass region where some $\Delta$ excitations predominate. 
Table~III shows, for baryons, the constituent quark mass, depending on the quark clusters considered. We observe the different cluster predictions for all five slopes.

For baryons, in the analysis of the narrow baryonic structures at 1.0$\le$M$\le$1.4~GeV \cite{bo1} \cite{bo2}, we have made the assumption that the clusters are 
 $(q{\overline q})^{2}-q^{3}$ with M$_{0}$=838.2~MeV and M$_{1}$=100.3~MeV. The first quantum numbers of the  
$(q{\overline q})^{2}-q^{3}$ configurations are the same that those of $q - q^{2}$ configurations, but these last are restricted to "classical" Particle Data Group (PDG) baryons
\cite{pdg}. Indeed there is no room for exotic baryons within these $q^{3}$ configurations \cite{caps}. Table III shows that in the mass range M$_{N}\le$M$\le$1.0~GeV, where the slope of the Regge-like trajectory equals 10.7~GeV$^{-2}$ such choice of configurations, namely $(q{\overline q})^{2}-q{^3}$ is indeed predicted.

In the mass range  M$_{N}\le$M$\le$1.0~GeV, following table III the masses of the narrow baryonic structures should be described with the lower mass cluster configurations. Following the previous remark, we use 
$q^{3} - q{\overline q}$ configurations to predict the masses.
\begin{figure}[h]
                                                       
\scalebox{.48}[.54]{
\includegraphics[bb=12 35 520 485,clip=]{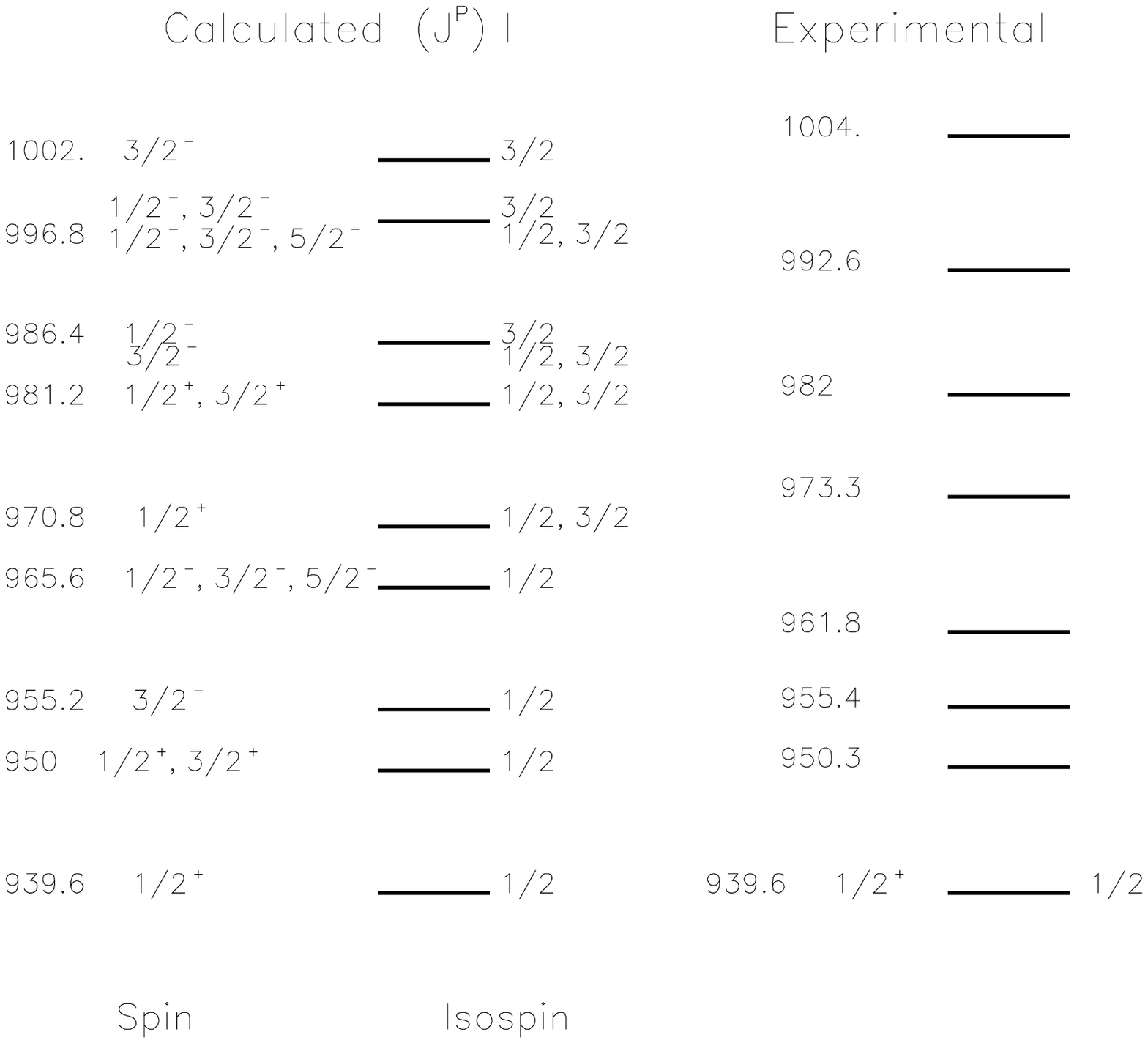}}
\caption{Experimental and calculated exotic baryonic masses in the range 945$\le$M$\le$1000~MeV.}
\label{fig17}

\end{figure}
The two parameters, M$_{0}$=924~MeV,
and M$_{1}$=15.6 MeV are adjusted in order to describe the neutron mass and
the first exotic mass at 950 MeV. The other masses, possible spins and
isospins, obtained with the use of formula [2], are shown in Fig.~17.
We observe the good
correspondance between the experimental and the calculated masses
since such correspondance is obtained without any other adjustable parameter.

\section{Dibaryons}
The spectrum of narrow exotic dibaryons was reported in \cite{bt}, where cross-sections of experiments performed at Saturne (SPES1 and SPES3) and in different other laboratories were reanalyzed. In the mass range studied, these masses are: M=1902, 1916, 1941, 1969, 2016, 2052, 2087, 2122, 2155, 2194, 2236, and 2282~MeV. 
These masses were reproduced quite satisfactory, using the mass formula [3] for $q^{4}-q^{2}$ quark clusters. Such agreement was obtained with $M_{0}$=1841~MeV and $M_{1}$=52.5~MeV.  Although the Barut's formula suggests more heavy clusters, with one or two additionnal $q{\overline q}$ contributions, the low calculated masses with 0 or 1 spin and isospin values are obtained indifferently of these assumptions for clusters.   

Notice that the same masses and quantum numbers are obtained if $q^{6} - q{\bar q}$
quark clusters will be considered. Indeed this last choice will allow additionnal masses  only above M=2681~MeV, therefore larger than those discussed here.

The corresponding Regge-like trajectory is shown in fig.~18.
\begin{figure}[!h]
                                                       
\scalebox{.95}[.49]{
\includegraphics[bb=45 20 300 520,clip=]{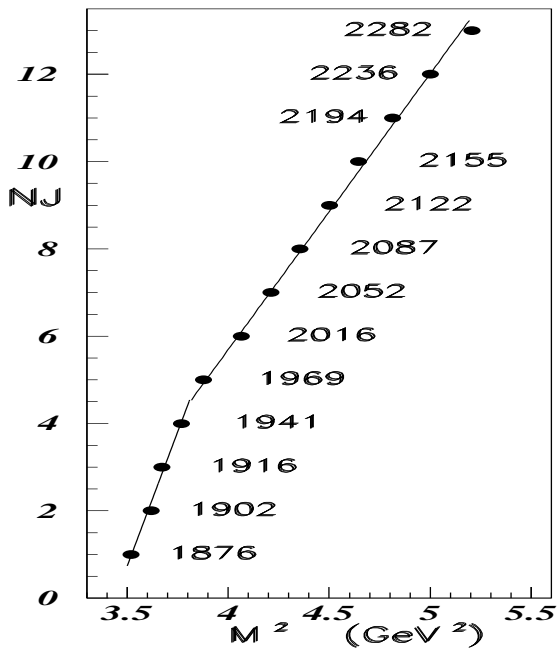}}
\caption{Regge-like trajectory of all narrow structure dibaryonic masses observed in \cite{bt}. The two slope values are: b=12.4~GeV$^{-2}$ and 6.3~GeV$^{-2}$.}
\label{fig18}

\end{figure}
\section{General discussion}
\subsection{Connection between exotic meson masses and exotic baryon masses}
When applying equation [2] for the calculation of exotic hadronic masses, we consider all possible spins and isospins obtained by addition of $q{\bar q}$ clusters to  $q{\bar q}$
for mesons and $q^{3}$ for baryons. It is therefore natural to compare the sequences of 
narrow mesonic and baryonic masses. The corresponding result is shown in fig.~19.

\begin{figure}[h]
                                                       
\scalebox{.5}[.47]{
\includegraphics[bb=20 20 530 530,clip=]{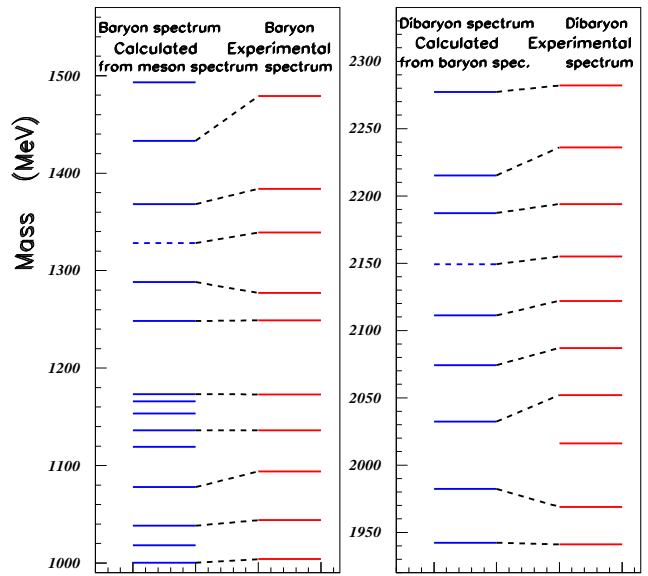}}
\caption{In the left side, comparison between the exotic baryonic mass spectrum obtained using the mesonic exotic masses, and the experimental one. In the right side, comparison between the experimental narrow dibaryonic spectrum with the calculated one starting from the experimental baryonic narrow mass spectrum.}
\label{fig19}
\end{figure}

The left part of the fig.~19 shows the masses of the experimental narrow baryons above 1~GeV compared to the calculated masses obtained by addition of the nucleon mass to the masses of narrow experimental mesons. The absence of some baryonic masses may be related to their experimental non observation due to poorer resolution than the one got for low mesonic mass measurements. The overall correspondance is noteworthy. 
The dashed calculated mass uses the mesonic experimental value (M=390~MeV), not extracted in previous papers, but extracted in this work - see section II.B.
The right part shows in the mass range 1920$\le$ M$_{dib.}\le$2300~Mev, the comparison between the experimental masses of narrow dibaryons and the masses obtained by addition of the nucleon mass to the narrow exotic baryonic masses. The dashed calculated mass, is obtained starting from an experimental mass M=1210~MeV, not extracted from the first analyses, but extracted \cite{pap} in the reanalysis of the $\gamma$p$\to\pi^{+}\pi^{0}$n reaction studied at MAMI \cite{lang} and the reanalysis of the H(e,e'$\pi^{\pm}$)X electroproduction cross-section measured at JLAB Hall C \cite{nav}.
\subsection{Discussion} 
For all three species, the slope (in GeV$^{-2}$) decreases for increasing masses. All extracted slopes are larger (in GeV$^{-2}$) than the slopes for "classical" PDG hadrons.

The importance of the concept of diquarks was often quoted in many papers. It was quantitatively discussed, on many examples of the PDG hadron spectroscopy by Sellem ans Wilczek \cite{sellem}. Mass formula [1] agrees with their statement "that the antisymmetric spin(isospin)-singlet state is more favorable energetically than the symmetric spin(isospin)-triplet state". Indeed, the calculated  masses increase with spin (isospin). 
 For baryons, the two first (3/2, 3/2) calculated masses have M=1200~MeV and M=1275~MeV. In \cite{bo4} it was shown that "the broad PDG baryonic resonances are in fact collective states of several weakly excited and narrow resonances." Then the mean calculated mass value between M=1200~MeV and M=1275~MeV is at 
$\Delta$M$_{c}$=299~MeV from the nucleon mass, to compare to the experimental value: $\Delta$M$_{e}$=290~MeV.

Following \cite{sellem} good diquarks have spin 0 and isospin 0. We apply it for our diquark clusters for mesons, and observe that such condition is fullfilled for the calculated masses: M=60~MeV, and is possible for M=100~MeV, 180~MeV, and 220~MeV. All three unobserved masses at M=120~MeV, 140~MeV, and 160~MeV, correspond to bad diquark cluster configurations (triplet spin (isospin) cluster state). 

\section{Conclusion}
In conclusion we have confirmed the existence of narrow structures in all three low mass hadronic species. The corresponding masses are reproduced with a small number of parameters, with help of a phenomenological mass relation [2]. These masses fit strait lines in Regge-like trajectories. This study stress on the importance of careful scrutiny of low mass hadronic spectroscopies.
\begin{table}
\vspace{5.mm}
\begin{tabular}{c c c c c c}
\hline
b (GeV$^{-2}$) & $q-q^{2}$&$q^{3}-q{\overline q}$&$q^{3}-
(q\overline q)^2 $&M$_{min}$&M$_{max}$\\
\hline
100.4  &{\bf 292.8} &168.6     &119.2      &939.6 &957 \\
62.6   &{\bf 369.7} &213.5     &150.9      &957   &1004\\
10.7   &893.7      &516       &{\bf 364.9} &1004  &1480\\
27.3   &559.7      &{\bf 323.2}&228.5      &1480  &1575\\
15.9   &733.8      &423.7     &{\bf 299.6} &1575  &1780\\
\hline
\end{tabular}
\caption{Constituent quark mass, for different clusters supposed to describe the baryonic narrow structures, derived from the Barut's equation \protect\cite{barut}. M$_{min}$ and M$_{max}$ give, in MeV, the mass range of each trajectory.}
\label{Table III}
\end{table}

We thank prof.  E.A. Kuraev for valuable discussions.

\end{document}